# CRIAÇÃO E APLICAÇÃO DE FERRAMENTA PARA AUXILIAR NO ENSINO DE ALGORITMOS E PROGRAMAÇÃO DE COMPUTADORES

*Afonso Henriques Fontes Neto Segundo – afonsof@unifor.br*
*Professor na Universidade de Fortaleza (UNIFOR)*
*Endereço: Av. Washington Soares, 1321 – Edison Queiroz, Fortaleza – CE, Brasil*

*Joel Sotero da Cunha Neto – joelsotero@unifor.br*
*Professor na Universidade de Fortaleza (UNIFOR)*
*Endereço: Av. Washington Soares, 1321 – Edison Queiroz, Fortaleza – CE, Brasil*

*Maria Daniela Santabaia Cavalcanti – mdaniela@unifor.br*
*Professora na Universidade de Fortaleza (UNIFOR)*
*Endereço: Av. Washington Soares, 1321 – Edison Queiroz, Fortaleza – CE, Brasil*

*Paulo Cirillo Souza Barbosa – pauloc@unifor.br*
*Pesquisador na Universidade de Fortaleza (UNIFOR)*
*Endereço: Av. Washington Soares, 1321 – Edison Queiroz, Fortaleza – CE, Brasil*

*Raul Fontenele Santana – raulfontenele@edu.unifor.br*
*Pesquisador na Universidade de Fortaleza (UNIFOR)*
*Endereço: Av. Washington Soares, 1321 – Edison Queiroz, Fortaleza – CE, Brasil*

**Resumo:** *O conhecimento sobre programação faz parte da matriz de conhecimentos que serão exigidos dos profissionais do futuro. Com base nisso, esse trabalho visa relatar o desenvolvimento de uma ferramenta de ensino elaborada durante o programa de monitoria da disciplina de Algoritmo e Programação de Computadores da Universidade de Fortaleza. A ferramenta une o conhecimento adquirido nos livros, com uma linguagem mais próxima aos alunos, utilizando vídeo aulas e exercícios propostos, com todo o conteúdo disponível na internet. Os resultados preliminares se mostraram positivos, com os alunos aprovando essa nova abordagem e acreditando que ela possa contribuir para um melhor desempenho na disciplina.*
**Palavras-chave:** Ferramenta de ensino. Programação. Internet.

## 1 INTRODUÇÃO

O modelo de ensino tradicional adotado na maioria das escolas brasileiras atualmente foi criado no Estados Unidos no início do século XX baseado no modelo de padronização e produção em massa utilizado pelas indústrias da época. O modelo separava os estudantes em séries com em suas idades e buscava agrupar alunos com o mesmo nível de conhecimentos em uma mesma sala para que o professor pudesse explicar a matéria de forma geral a todos. Esse modelo de ensino apresentava bons resultados visto que no período os empregos, em sua maioria, não exigiam um elevado nível de conhecimento por parte dos funcionários. Com os avanços da tecnologia houve mudanças no mercado de trabalho e atualmente cerca de 60% dos empregos exigem trabalhadores com um elevando grau de instrução, o que tornou esse modelo de ensino insuficiente. No modelo tradicional o professor explica o conteúdo uma vez, e devido às diferentes velocidades de aprendizado, alguns alunos acabam não aprendendo a matéria de forma imediata, criando uma lacuna que irá dificultar o aprendizado da matéria subsequente (HORN; STAKER; CHRISTENSEN, 2015).

Como indicado em Kazakoff, Sullivan, Bers (2013), a programação é uma importante habilidade na alfabetização no século XXI e que será necessária para a maioria das carreiras futuras. Essa importância pode ser vista em trabalho como Rebouças et al (2010) ou Grandell et al (2006) nos quais são apresentadas e discutidas propostas para o ensino de programação em escolas através do desenvolvimento de jogos com a linguagem Python. Jenkins (2002) destaca que o ensino de programação pode se tornar uma tarefa complicada, pois não exige apenas que o aluno conheça as regras de sintaxe da linguagem que está usando, exige também que o aluno possua raciocínio lógico e capacidade de abstração. Além das habilidades já mencionadas, um fator primordial para a resolução de problemas ligados a programação é a experiência, o que significa que é para um melhor aprendizado de programação é necessário programar (JENKINS, 2002).

Gomes (2008) reforça ainda que o ensino das linguagens de programação possuem objetivos que vão além do simples aprendizado da sintaxe da língua, visando proporcionar ao aluno um conjunto de competências que o levem a solucionar problemas do cotidiano, demandando dos estudantes muito esforço e perseverança, como também um grande poder de abstração para que os mesmos consigam resolver problemas de forma genérica (GOMES, 2008).

Uma nova perspectiva para auxiliar o aluno nesse desenvolvimento é o ensino on-line, o qual começou visando atender estudantes em circunstâncias nas quais não havia outra alternativa para a aprendizagem, porém atualmente há um crescimento na quantidade de estudantes experimentando o ensino virtual e que continuam a frequentar escolas físicas tradicionais, fenômeno conhecido como ensino hibrido (HORN; STAKER; CHRISTENSEN, 2015).

O trabalho apresentado no presente artigo visa relatar o desenvolvimento de uma ferramenta on-line de ensino de programação capaz de se integrar à disciplina de Algoritmos e Programação da Universidade de Fortaleza (UNIFOR) e fornecer mais opções confiáveis para o estudo à distância aos alunos.

## 2  PÚBLICO ALVO

O público alvo do trabalho foram os alunos da disciplina de Algoritmos e Programação de Computadores da Universidade de Fortaleza (UNIFOR). Essa disciplina faz parte da matriz curricular obrigatória de todas as engenharias do centro de ciências tecnológicas da universidade e está alocada, em geral, no primeiro semestre.

A fim de entender o perfil dos alunos da disciplina foi elaborado um questionário, que foi aplicado em três turmas distintas. Para a escolha das turmas, foi levado em consideração a quantidade de alunos e o turno (manhã ou noite), já que foi constatado que, historicamente, as turmas dos turnos da manhã e da noite possuem perfis distintos. As turmas noturnas costumam possuir alunos com idades mais avançadas, que já estão realizando sua segunda graduação ou que possuem algum trabalho durante o dia, enquanto as turmas matutinas costumam ser formadas por alunos mais jovens, com idades entre 17 e 22 anos, que estão cursando sua primeira graduação e que não possuem um trabalho formal.

A primeira pergunta do questionário solicitava o curso no qual o estudante estava matriculado. Para essa pergunta foram obtidas como respostas todos os cursos de engenharia do centro de ciências tecnológicas, que abrange Engenharia Civil, Elétrica, Mecânica,

Eletrônica, Controle e Automação, Produção, Computação e Ambiental e Sanitária. Conforme o esperado, cerca de 40% dos entrevistados cursavam Engenharia Civil, que atualmente é o curso com a maior quantidade de alunos entre as engenharias, enquanto os outros cursos apresentaram parcelas aproximadamente iguais. Ainda foram obtivas três respostas para a opção "Outros", o que indica alunos fazendo a disciplina de forma optativa. Quando perguntado o semestre que no qual os mesmos estavam cursando, foram obtidas respostas que variavam do primeiro ao oitavo semestre, com cerca de 30% das respostas indicando o primeiro semestre. Quando questionados se os mesmos já haviam reprovado ou trancado a disciplina, 45% dos entrevistados responderam positivamente. Esses números, quando confrontados com o semestre no qual os alunos estão cursando, indicam que muitos alunos estão cursando a disciplina em semestres diferentes do sugerido pela matriz curricular. Quando questionados sobre quais ferramentas utilizavam para o estudo da disciplina, como mostrado na Figura 1, a opção "Lista de exercícios" e "Vídeo aulas" foram marcadas em 48% e 35% dos casos respectivamente. Ao final do questionário foi indagado a que ferramentas os mesmos recorriam em caso de dúvidas e 26% dos entrevistados indicaram recorrer à *internet* como principal opção, como apresentado na Figura 2. Ao total foram entrevistados 54 alunos.

**Figura 1 - Ferramentas mais utilizadas pelos alunos**

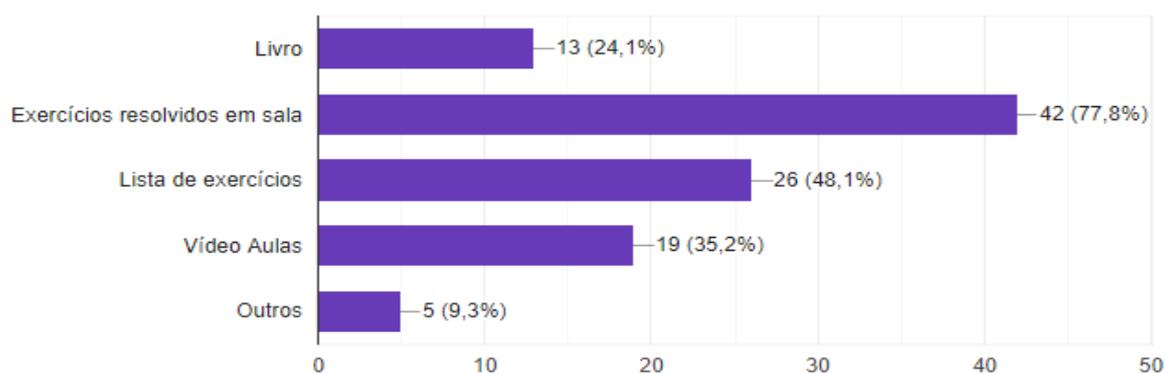

**Fonte: Autor**

**Figura 2 - Opção dos alunos em caso de dúvida**

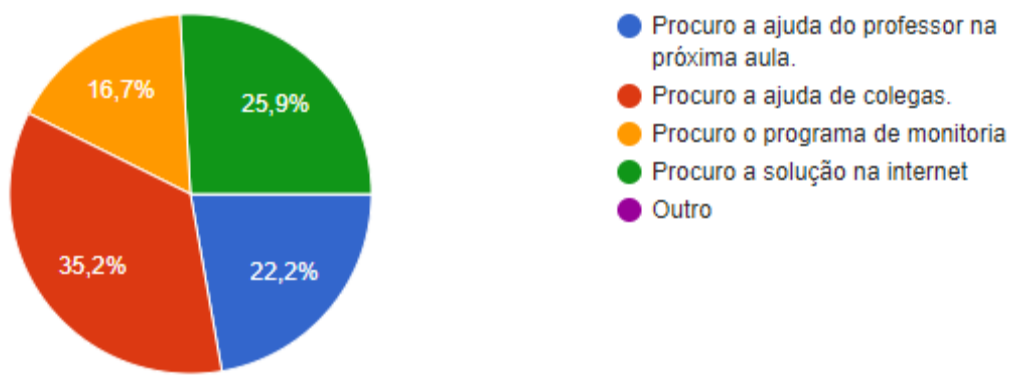

## 3 TRABALHOS RELACIONADOS

Em Machado et al (2018) é apresentado um aplicativo de celular para auxiliar no ensino de programação orientada a objetos. Esse aplicativo possui três opções de atividades. Na primeira atividade, são apresentadas questões de múltipla escolha que podem ser usadas para explorar o conteúdo conceitual da disciplina ou como forma de avaliação. Na segunda atividade é proposto uma espécie de quebra cabeças, onde o aluno deve ordenar blocos de código a fim de chegar em um algoritmo final correto. Na terceira, são propostos problemas abertos, nos quais os alunos devem enviar desde explicações até códigos realizados de forma livre. O aplicativo possui conectividade com a internet e permite uma troca de mensagens entre os alunos e o professor em seu interior. Ao final do trabalho o autor concluiu que apesar das limitações ligadas à necessidade dos alunos possuírem dispositivos capazes de se conectar com a internet, o trabalho traz melhoras para o aprendizado da disciplina, uma vez que houve uma queda significativa no índice de reprovação da disciplina em relação aos semestres anterior.

Em Verdu *et al*. (2012) é proposto uma ferramenta avaliativa que combina um servidor capaz de fornecer um feedback não binário sobre o problema enviado com um repositório de problemas que os alunos podem acessar de acordo com o seu perfil e suas habilidades. A ferramenta tenta fomentar o estudo de programação através de um sistema de feedback e de competições para que os alunos sempre se mantenham motivados a aprender cada vez mais. Ao fim do trabalho o autor chegou à conclusão que os alunos que usaram a ferramenta de ensino obtiveram melhores notas e com uma menor variação entre notas extremas, em relação aos alunos que não utilizaram.

## 4 DESENVOLVIMENTO DA FERRAMENTA

A ferramenta foi criada com o propósito não apenas de auxiliar os alunos no estudo da disciplina de Algoritmos e Programação de Computadores fora da sala de aula, mas desmistificar a lógica de programação, enfatizando importância desse conhecimento para a inserção em um mercado de trabalho cada vez mais exigente.

O preparo de todo o material que será apresentado foi desenvolvido por alunos e ex-alunos do programa de monitoria da Universidade de Fortaleza em conjunto com os professores da disciplina. Para que todo o material ficasse disponível na *internet*, como forma de atingir o maior público possível, foi usado o site GitHub, que é amplamente utilizado por programadores.

O GitHub é uma plataforma de hospedagem de códigos gratuita que permite o controle de versões de código e colaboração entre usuários usando repositórios. Os repositórios são partições utilizados para separarem projetos e podem conter imagens, vídeos, planilhas ou o que mais for necessário para a execução de um projeto (GitHub, 2019). O site possui a opção de criar repositórios públicos ou privados. Os *repositórios* classificados como públicos permitem que qualquer indivíduo que entre no *site*, possua cadastro ou não, possa realizar a visualização e o *download* das informações lá contidas, enquanto os privados permitem a visualização apenas dos usuários responsáveis ou que foram previamente autorizados a interagir com os arquivos.

O material começa com uma breve introdução sobre o que são algoritmos, qual sua importância, buscando mostrar o porquê esse tipo de conhecimento faz parte das habilidades que serão exigidas dos profissionais do futuro e quais suas aplicações no cotidiano. Para ratificar isso, é apresentado também uma videoaula que enfatiza a explicação anterior e apresenta exemplos simples de algoritmos e como é possível resolvê-los. Para o desenvolvimento do restante do conteúdo a ferramenta, inicialmente o conteúdo programático foi particionado em 7 tópicos principais, que são *input/output, if/else, while, for,* funções, vetores e matrizes, e cada tópico foi dividido em 3 seções, que seguiram o mesmo padrão em cada tópico. Os tópicos estavam distribuídos e apresentados na mesma ordem citada anteriormente e seguiam a ordem da apresentação dos conteúdos em sala de aula.

A primeira seção de cada tópico possuía uma breve explicação sobre o comando e suas funcionalidades. A explicação visa unir o conteúdo presente nos livros de programação, com uma linguagem simples, clara e direta que pudesse motivar a leitura por parte dos alunos. Seguida a explicação era apresentada, em forma de código, a estrutura básica do comando e como utilizá-lo. Na figura 3 é apresentada a estruturação da função *if,* que é amplamente usada como estrutura de seleção baseada em condição.. Essa estrutura permite que certa parte do código só seja executada caso uma determinada condição seja atendida.

Na segunda seção, é apresentado um problema proposto, a escolha do problema se daria de forma que o escolhido pudesse ser uma aplicação simples do comando aprendido no tópico atual e, devido o conteúdo ser acumulativo, poderia exigir conhecimentos adquiridos em tópicos anteriores. Juntamente com cada um dos problemas propostos é apresentado uma videoaula, feita pelos autores, com a explicação do comando e de como resolver o desafio apresentado. Durante a resolução do problema, os autores buscavam sempre destacar os erros mais comuns cometidos pelos alunos que puderam ser catalogados ao longo dos dois anos e meios em que os autores participaram do programa de monitoria. Também é disponibilizado para o aluno a opção de realizar o *download* do código fonte utilizado na aula para que o mesmo possa executar o *script* e a partir dele realizar seus próprios testes.

Na terceira seção de cada tópico é apresentada uma lista de questões para estudo. A escolha das questões se deu através de materiais e livros didáticos disponibilizados e adotados pelos professores durante a disciplina, e sua disposição foi planejada para que fossem apresentadas em uma ordem crescente de dificuldade, de forma que o fato do estudante ter realizado a questão anterior facilitaria no desenvolvimento da questão subsequente.

Figura 3 - Estruturação do if

```
if (primeira_condição_a_ser_testada)
{
        bloco de código para o caso no qual a condição testada seja verdadeira;
}
else if(segunda_condição_a_ser_testada)
{
        Bloco de código que será executado caso a condição dentro do else if seja verdadeira;
        //Não existe a obrigatoriedade de existir, mas o mesmo só pode existir após um if.
}
else{
        Bloco de código que será caso nem a primeira nem a segunda condição sejam verdadeiras;
        //Não existe a obrigatoriedade de existir, mas o mesmo só pode existir após um if.
}
```

Fonte: Autor

## 5 TRABALHOS FUTUROS

O atual trabalho possui as ideias de um repositório e um banco de questões classificadas por nível de dificuldade e conhecimento apresentado em Verdu et al. (2012), porém ainda deixa a desejar em relação ao *feedback* dos trabalhos desenvolvidos pelos alunos, uma vez que os mesmo só os tem durante os horários disponíveis para a monitoria ou em sala de aula com o professor. Para corrigir tal limitação, o próximo passo do trabalho é a integração da ferramenta atual com outras ferramentas de ensino que possibilitem um retorno mais instantâneo aos alunos. Outro ponto a ser melhorado no futuro é uma maior integração da atual ferramenta com o sistema de comunicação interna da universidade, que permite a criação de grupos para que os alunos possam discutir e se ajudar em um determinado assunto, como o ambiente apresentado em Machado et al (2018). Também foi planejada a inclusão de novas questões e um aumento em seus níveis de dificuldade que permitam que os alunos se aprofundem no assunto.

## 6    CONSIDERAÇÕES FINAIS

A fim de descobrir a impressões dos estudantes após a utilização da ferramenta, no final da página do GitHub foi apresentado um segundo questionário. O questionário foi dividido em 3 partes, com perguntas direcionadas aos aspectos organizacionais, abordagem e conteúdo.

A primeira parte do questionário buscava saber grau de utilizada que os alunos atribuíam a ferramenta e o quão intuitivo tinha sido o seu uso. As opções estavam graduadas de 1 a 5, onde 1 representava muito pouco e 5 representava bastante. Quando perguntados sobre a utilidade, as respostas 4 e 5 apresentaram 38,7% e 48,4% das alternativas marcadas respectivamente. Para a facilidade do uso as respostas 4 e 5 apresentaram 32,3% e 58,1% das alternativas marcadas respectivamente.

A segunda parte buscava saber a percepção dos alunos sobre as ferramentas auxiliares usadas. Quando perguntado sobre o grau de relevância das videoaulas contidas na ferramenta, com a mesma gradação usada nas perguntas anteriores, as alternativas 4 e 5 apresentaram 30% e 60% das respostas marcadas respectivamente, o que pode caracterizar uma ótima aceitação desse tipo de abordagem. Ao serem questionados a respeito dos exercícios resolvidos, cerca de 97% dos entrevistados afirmaram que os tópicos foram desenvolvidos de forma clara e direta, mostrando que este objetivo foi atingido.

A terceira parte do questionário buscava saber se o material desenvolvido havia atingido seu objetivo de melhorar o desempenho dos alunos na disciplina. Para isso foi perguntado se os mesmos acreditavam que a ferramenta poderia ajudá-los a aumentar seu rendimento e a resposta "sim" foi unânime. Devido ao alto índice de reprovações e trancamentos, foi perguntado se os alunos acreditavam que esse material poderia ajudado a evitar tal situação e, conforme demonstrado na Figura 4, a resposta "sim" representou 45,2% das respostas, considerando que 48,4% afirmaram estar fazendo pela primeira vez. Por fim foi questionado se os mesmos indicariam a ferramenta para algum amigo que ainda não cursou a disciplina e que gostaria de aprender a programar e, conforme mostrado na figura 5, a resposta positiva foi marcada em 93,5% dos casos.

**Figura 4 - Resposta referentes ao questionado sobre evitar reprovação ou trancamento**

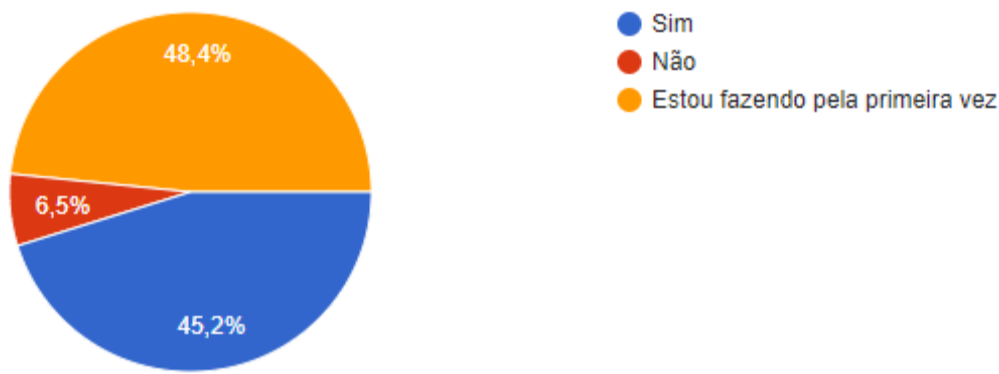

**Fonte: Autor**

**Figura 5 - Respostas sobre a indicação da ferramenta**

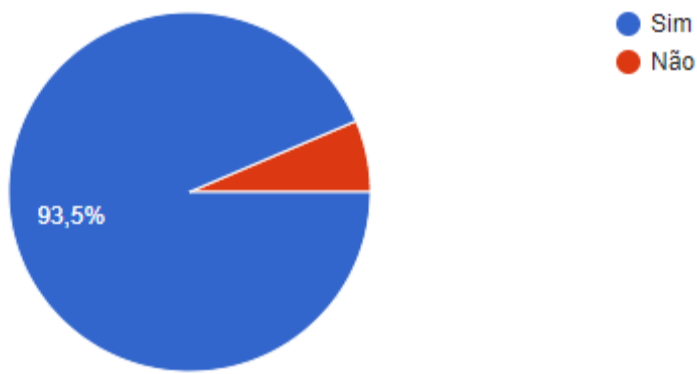

**Fonte: Autor**

As respostas do segundo questionário mostraram que o objetivo inicial do trabalho foi alcançado, o que traz uma maior motivação para a continuidade do desenvolvimento de um material cada vez mais completo e que possa alinhar o ensino de Algoritmos e Programação de Computadores com as ferramentas mais modernas já desenvolvidas.

# 7 REFERÊNCIAS BIBLIOGRÁFICAS

https://guides.github.com/activities/hello-world/ - site do github – 23/04 as 00:50

HORN, Michael B.; STAKER, Heather; CHRISTENSEN, Clayton. **Blended: usando a inovação disruptiva para aprimorar a educação**. Penso Editora, 2015.

KAZAKOFF, Elizabeth R.; SULLIVAN, Amanda; BERS, Marina U. The effect of a classroom-based intensive robotics and programming workshop on sequencing ability in early childhood. **Early Childhood Education Journal**, v. 41, n. 4, p. 245-255, 2013.

REBOUÇAS, Ayla Débora Dantas S. et al. Aprendendo a ensinar programação combinando jogos e Python. In: **Brazilian Symposium on Computers in Education (Simpósio Brasileiro de Informática na Educação-SBIE)**. 2010.

GRANDELL, Linda et al. Why complicate things?: introducing programming in high school using Python. In: **Proceedings of the 8th Australasian Conference on Computing Education-Volume 52**. Australian Computer Society, Inc., 2006. p. 71-80.

GOMES, Anabela et al. Aprendizagem de programação de computadores: dificuldades e ferramentas de suporte. **Revista Portuguesa de Pedagogia**, p. 161-179, 2008.

JENKINS, Tony. On the difficulty of learning to program. In: **Proceedings of the 3rd Annual Conference of the LTSN Centre for Information and Computer Sciences**. 2002. p. 53-58.
GOMES, Anabela et al. Aprendizagem de programação de computadores: dificuldades e ferramentas de suporte. **Revista Portuguesa de Pedagogia**, p. 161-179, 2008.

MACHADO, Leonardo Davi Pereira et al. Uma ferramenta colaborativa para apoiar a aprendizagem de programação de computadores. **Revista Brasileira de Computação Aplicada**, v. 10, n. 1, p. 23-29, 2018.

VERDÚ, Elena et al. A distributed system for learning programming on-line. **Computers & Education**, v. 58, n. 1, p. 1-10, 2012.

# CREATION AND APPLICATION OF A TOOL FOR AUXILIARY IN ALGORITHM EDUCATION AND COMPUTER PROGRAMMING

*Abstract:* *Knowledge about programming is part of the knowledge matrix that will be required of the professionals of the future. Based on this, this work aims to report the development of a teaching tool developed during the monitoring program of the Algorithm and Computer Programming discipline of the University of Fortaleza. The tool combines the knowledge acquired in the books, with a language closer to the students, using video lessons and exercises proposed, with all the content available on the internet. The preliminary results were positive, with the students approving this new approach and believing that it could contribute to a better performance in the discipline.*

*Key-words:* *Teaching tool, programming, internet.*